# Backscattering-Based Security in Wireless Power Transfer Applied to Battery-Free BLE Sensors

Taki Eddine DJIDJEKH
LAAS-CNRS, Université de Toulouse,
CNRS, Ogoxe
Toulouse, France
taki-eddine.djidjekh@laas.fr

Gaël LOUBET
LAAS-CNRS, Université de Toulouse,
CNRS, INSA
Toulouse, France
gael.loubet@laas.fr

Alexandru TAKACS
LAAS-CNRS, Université de Toulouse,
CNRS, UPS
Toulouse, France
alexandru.takacs@laas.fr

*Abstract*— The integration of security and energy efficiency in Internet of Things systems remains a critical challenge, particularly for battery-free and resource-constrained devices. This paper explores the scalability and protocol-agnostic nature of a backscattering-based security mechanism by integrating it into Bluetooth Low Energy battery-free Wireless Sensor Network. The proposed approach leverages the Wireless Power Transfer link, traditionally used for energy harvesting, to generate additional identification signals without increasing energy consumption or computational demands. Experimental validation demonstrates the solution's functionality using compact, low-gain antenna, ensuring compatibility with size-constrained applications such as Structural Health Monitoring and smart transport. Furthermore, this work addresses the challenges associated with backscattering dynamic range and multi-node Wireless Sensor Network scenarios, discussing potential collisions between identification signals and proposing future improvements to enhance generalizability and scalability. The findings underscore the potential of the backscattering-based security mechanism for creating secure, sustainable, and scalable IoT deployments across diverse protocols and applications.

I. INTRODUCTION

The Internet of Things (IoT) is rapidly advancing, with an increasing deployment of wireless nodes operating in Cyber-Physical Systems (CPS) across diverse applications, including Structural Health Monitoring (SHM), biomedical devices, and autonomous systems. These Wireless Sensor Networks (WSNs) face two critical challenges: ensuring robust security and achieving sustainability through energy efficiency and autonomy. As the demand for scalable, energy-efficient IoT solutions grows, addressing the interplay between security and energy autonomy becomes paramount, especially for battery-free IoT nodes.

Traditional IoT security relies on software cryptographic methods to mitigate threats such as unauthorized access, replay attacks, and eavesdropping. While effective, these methods often impose significant computational and energy demands, making them unsuitable for energy-constrained and battery-free systems [1]. Concurrently, energy efficiency is being pursued through energy harvesting, Wireless Power Transfer (WPT) [2], and lightweight communication protocols like LoRaWAN and Bluetooth Low Energy (BLE) [3]. However, integrating security into sustainable, battery-free IoT systems remains a complex challenge, especially in the case of Simultaneous Wireless Information and Power Transfer (SWIPT).

Several approaches have been explored to address the dual challenges of energy efficiency and security in SWIPT systems. For instance, Full-Duplex Jamming with Power Splitting combines energy harvesting with interference generation to prevent eavesdropping, though it requires additional hardware for managing interferences [4]. Similarly, Intelligent Reflecting Surfaces (IRS) dynamically direct signals to legitimate receivers, enhancing security but at the cost of sophisticated hardware, limiting scalability in lightweight IoT deployments [5]. Direct Spread Spectrum Techniques (DSSS) provide security when integrated with SWIPT, but their hardware complexity restricts their applicability to DSSS-specific IoT standards [6]. Other physical-layer methods, such as robust beamforming and artificial noise generation, improve secrecy rates and energy efficiency under imperfect channel conditions but demand complex algorithms [7]. Despite these advancements, the reliance on intricate hardware and algorithms often limits their feasibility in resource-constrained IoT environments.

A promising solution to address the trade-off between security and energy efficiency is the backscattering-based security mechanism introduced in [8]. This approach utilizes a backscattering rectifier to provide lightweight, energy-efficient security by leveraging the WPT link, traditionally used for energy harvesting, to generate additional identification signals. This process operates independently of the IoT protocol data link, requiring no additional energy consumption, computational resources, or complex algorithms. Its integration into LoRaWAN networks demonstrated both functionality and feasibility in [9], validating the approach within a specific communication protocol.

This paper scales the backscattering solution to BLE-enabled battery-free nodes, showcasing its protocol-agnostic scalability and its potential for broader IoT applications. The main contributions include the integration of the backscattering rectifier into BLE-based WSNs, demonstrating its applicability beyond LoRaWAN. Additionally, the study demonstrates a sustainable security mechanism tailored for battery-free systems, minimizing energy consumption while maintaining robust functionality. Experimental validation is conducted using compact, low-gain antenna to prove the solution's effectiveness and compatibility with size-constrained applications, such as autonomous transport systems and Structural Health Monitoring. The paper also discusses the generalizability of the solution in multi-node WSNs, addressing challenges such as avoiding collisions between identification signals from different nodes. Finally, it explores the broader implications, limitations, and potential future improvements to enhance scalability and effectiveness in diverse IoT scenarios.

The remainder of this paper is structured as follows: Section II reviews the backscattering security technique and its integration into BLE-based Wireless Sensor Networks. Section III presents the experimental setup and validation process. Section IV discusses the results, highlights limitations, and explores directions for future improvements. Finally, Section V concludes with a summary of the contributions and their implications for secure, sustainable IoT deployments.

## II. BACKSCATTERING-BASED SECURITY MECHANISM AND BLE INTEGRATION

### A. WPT-Based Backscattering Security Mechanism

The backscattering-based security mechanism effectively addresses the dual challenges of security and energy efficiency in SWIPT systems by leveraging the WPT link, traditionally used exclusively for energy harvesting, to generate an additional uplink signal for secure identification. This approach operates independently of the data link, ensuring compatibility with existing lightweight IoT protocols. Fig. 1 illustrates the architecture of a SWIPT system enhanced with the WPT-based backscattering identification mechanism.

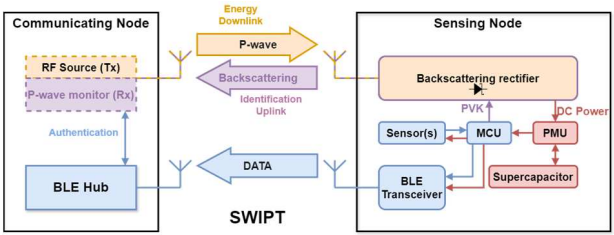

Fig. 1. Architecture of SWIPT system enhanced with WPT-based backscattering security mechanism.

On the battery-free Sensing Node (SN) side, a backscattering rectifier is integrated, designed to be modulated by a digital signal with high and low states, enabling the dynamic modulation of the backscattered power wave (P-wave). On the Communicating Node (CN) side, a P-wave monitor is introduced to extract the security identification encoded within the modulated P-wave, ensuring a robust and energy efficient authentication and security mechanism. The data link remains the same as in the traditional battery-free SWIPT architecture [10].

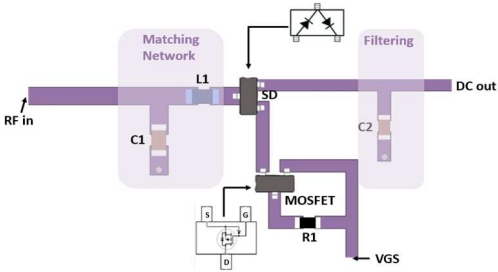

Fig. 2. Backscattering Rectifier circuit.

The Backscattering Rectifier (BR) is a modified RF rectifier designed for dual functionality: energy harvesting and backscattering. Using minimal components, including a MOSFET (BSS123), an LC matching network and a Schottky diode pair (SMS7630-005LF), it efficiently converts RF energy into dc power when VGS = 0V and enables backscattering when VGS = 3.3V. The layout and components are illustrated in Fig. 2.

The SN leverages a General Purpose Input/Output (GPIO) digital pin from its low-power microcontroller (MCU) to control the BR and encode an identification sequence as a Private Key (PvK), eliminating the need for high computational capabilities or significant energy consumption. This process can be performed while the transceiver remains inactive. Once the identification is sent, the SN resumes its standard operations, such as measurement and data transmission, without any changes to its usual functionality. On its side, the CN verifies the backscattered sequence using the P-wave monitor to determine whether to accept or reject the data. This mechanism operates in addition to the lightweight standard protocol security implemented at the data link level, providing an extra layer of authentication and security.

### B. Integration into BLE-Based Wireless Sensor Network

An autonomous Battery-Free Wireless Sensing Node (BFSN) operating with BLE for data transmission and radiative WPT [11] was chosen for integration with the BR to test the WPT-based backscattering security mechanism in BLE WSN. This BFSN, uses a 220 µF storage capacitor, was optimized for performing sensing tasks (temperature and humidity measurements) and BLE broadcasting with low energy requirements, every time its storage is charged by radiative harvesting (cycle).

To integrate the BR, the BFSN's standard rectifier was bypassed, and the BR's DC output was connected to the input of the Power Management Unit (PMU). A GPIO pin from its QN9080 NXP chip was linked to the BR's VGS input, enabling control of the backscattering mechanism. On the software side, minor modifications were implemented to generate a Manchester-encoded identification sequence by toggling the GPIO pin. This sequence is transmitted during brief backscattering events every cycle, while regular sensing and BLE broadcasting tasks resumed seamlessly after that. The BR integrated into the BFSN is illustrated in Fig. 3.

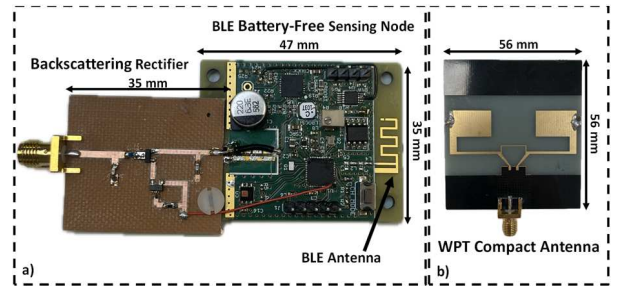

Fig. 3. a) Backscattering Rectifier integrated into BLE battery-free wireless sensing node; b) Compact 868MHz Antenna.

A compact, low-gain dipole antenna with dimensions 56 mm × 56 mm × 10 mm and a gain of +1.1 dBi was utilized to ensure compatibility with space-constrained applications and evaluate effectiveness of the mechanism. This antenna optimized for operation in the ISM 868 MHz band, balancing compactness and performance [12].

## III. EXPERIMENTS AND RESULTS

The embedded software modifications were first tested using a portable Oscilloscope (PicoScope 2205) to verify the correct generation of the Manchester-encoded PvK sequence. The QN9080 chip of the BLE BFSN operates at a voltage of 3V, resulting in a high GPIO output of 3V. While the optimal reflection for the Backscattering Rectifier is achieved with VGS=3.3V, a slight reduction in reflection performance was expected due to this voltage difference.

Initially, the PvK sequence was generated at a frequency of 100kHz, providing a trade-off between shorter sequence duration (approximately 2.5 ms for 16 bytes) to minimize energy consumption and the switching capability of the MOSFET (BSS123) without signal distortion. However, the GPIO pins of the QN9080 have limited switching speeds, constraining the sequence generation to a maximum frequency of 40 kHz to avoid distortion. At this frequency, the sequence duration increased to approximately 7ms. Fig. 4. illustrates the Manchester-encoded PvK sequence as measured with the PicoScope at a frequency of 40 kHz.

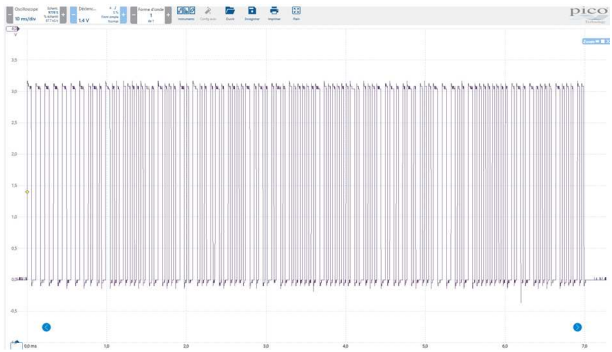

Fig. 4. PicoScope screenshot showing the Manchester-encoded PvK sequence measured on the BLE sensing node's GPIO pin.

For the wireless setup of the BLE WSN, testing was conducted outside the anechoic chamber to evaluate performance under real-world conditions. The BLE BFSN was positioned 1.3 meters away from the CN. The setup utilized a patch antenna with a gain of +9.2 dBi at 868 MHz, connected to the RF source via an RF circulator.

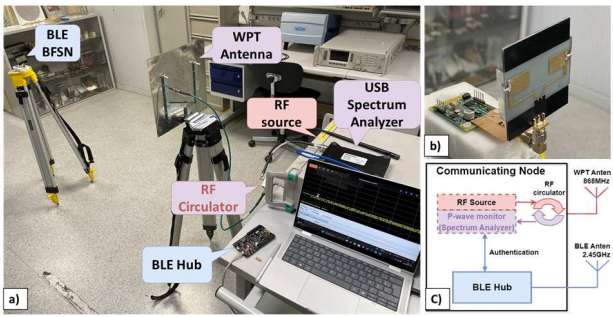

Fig. 5. Experimental Setup: a) Photograph of the entire setup; b) Photograph of the Battery-Free Sensing Node in the setup; c) Schematic of the Comunicating Node setup.

An Anritsu MG3694B RF signal generator served as the RF source, delivering the P-wave through the RF circulator (Aerotek C11-1FFF). The circulator's second port was connected to the +9.2 dBi patch antenna to emit the P-wave and receive the backscattered signal, while its third port was linked to a USB Spectrum Analyzer (Tektronix RSA306B) for monitoring the backscattered signal as the P-wave Monitor.

The RF circulator provided 20 dB of isolation between the RF source port and the Spectrum Analyzer port, Fig. 5. illustrates the setup. NXP development card (QN9080-DK) was used as BLE Hub to retrieve sensor measurement data.

At an RF source output power level of 18 dBm at 868MHz, the BLE BFSN successfully performed both backscattering of the sequence and transmission of measurement data. The system operated in cycles, with the storage element charging every 10 seconds. Both the data and the backscattered sequence were correctly received. The backscattered sequence observed on the Spectrum Analyzer is illustrated in Fig. 6.

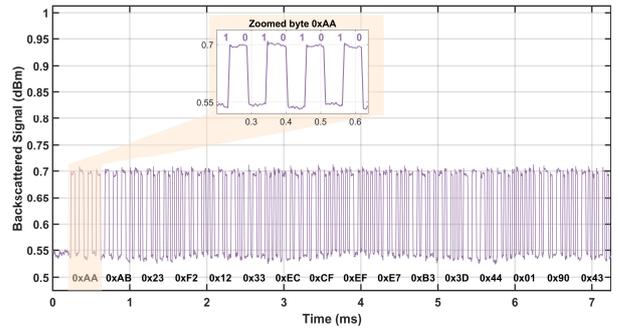

Fig. 6. Backscattered signal (16 bytes identification Manchester-encoded sequence) captured at the spectrum analyzer with a zoom in the first byte.

Although the signal was captured by the Spectrum Analyzer using triggering, the dynamic range (difference between the high and low states) of the backscattered signal was relatively low, approximately 0.15dB. This reduced dynamic range can be attributed to the slightly lower reflection caused by the 3V maximum high output of the BFSN chip, the low-gain antenna, and environmental factors. Additionally, a high level of cross-jamming, resulting from environmental reflections, and the limited isolation of the RF circulator between the emission and reception ports, further contributed to signal degradation, along with propagation losses.

Despite these challenges, the backscattering security technique was demonstrated to function within the BLE WSN setup, underscoring its scalability and adaptability to IoT applications across various communication protocols.

## IV. DISCUSSION

The experimental results of the backscattering-based security mechanism integrated into a BLE battery-free sensing node demonstrate its feasibility and adaptability to real-world IoT applications. However, the findings also highlight several practical challenges and considerations that merit further discussion.

### A. Functionality and Performance

The backscattering security mechanism successfully operated within the BLE WSN, enabling both secure identification and data transmission. At an RF source output power of 18 dBm at 868 MHz, the BFSN performed its tasks, including charging, backscattering the Manchester-encoded sequence, and transmitting measurement data. The backscattered sequence and the data were correctly received, confirming the functionality of the system and demonstrating its scalability and protocol-agnostic nature. The ability to integrate this mechanism into BLE, in addition to its prior implementation in LoRaWAN networks, underscores its adaptability to a wide range of IoT protocols. This scalability highlights the potential for broader IoT deployments without significant modifications to existing infrastructure.

*B. Signal Dynamics and Challenges*

While the backscattered sequence was detected using a Spectrum Analyzer, the dynamic range of the signal was limited to approximately 0.15 dB. Several factors contributed to this:

- Voltage output of the BFSN chip: The 3V high output of the QN9080 chip, compared to the optimal VGS of 3.3V for the Backscattering Rectifier, slightly reduced the efficiency of the backscattering.

- Antenna Gain: The use of a compact low-gain antenna further limited the signal strength, though its compactness is essential for integration into space-constrained IoT devices.

- Cross-Jamming: Reflections and environmental noise created high levels of cross-jamming, complicating signal detection. The RF circulator provided only 20 dB of isolation between the emission and reception ports, which, coupled with propagation losses and clutter effects, further weakened the backscattered signal and increased the cross-jamming.

*C. Future Improvements*

To address the identified challenges and improve performance:

- Enhanced Voltage Matching: Employing impedance matching of the BR to achieve maximum reflection in function of the IoT device voltage.

- Higher Isolation Circulator: Using RF circulators with greater isolation between ports would reduce cross-jamming and enhance signal clarity.

- Optimized Antenna Design: Exploring compact antenna designs with slightly higher gain could balance size constraints with improved signal dynamics.

Furthermore, the current study did not evaluate scenarios with multiple BFSNs sharing the same communicating node. In such cases, simultaneous backscattered identification sequences could interfere with one another. To address this, further tests are needed to assess interference probabilities. Nevertheless, this self-interference has a very low probability because of the inherent battery-free operation mode of the BFSN driven by the charging duty cycle of the energy storage elements that is strongly dependent of the BFSN location. Also, this (self-interference) probability can be further minimized by controlling the timestamp of the backscattering frames. Another solution is to adapt the BRs of different BFSNs to operate within different tightly defined frequency bandwidths. This approach would allow the RF source to emit a broader signal while limiting the backscattered sequences of each BFSN to distinct frequency bands, effectively reducing interference, and adding an additional security.

## V. Conclusion

This paper demonstrated the successful implementation of a backscattering-based security mechanism in the context of Simultaneous Wireless Information and Power Transfer for a battery-free Bluetooth Low Energy wireless sensor. By integrating a backscattering rectifier and utilizing the energy wave used for power transfer, a hardware-based security solution was introduced that is complementary to and fully interoperable with the BLE protocol.

This mechanism provides a lightweight and energy-efficient security layer that enhances existing protocol-level measures, making it particularly suitable for resource-constrained IoT applications. The experimental validation confirmed the system's functionality and feasibility under real-world conditions, demonstrating its adaptability to BLE-based IoT networks. By successfully applying this concept to BLE, this study advances secure and sustainable IoT deployments. The findings highlight the potential for broader adoption of backscattering-based security mechanisms, paving the way for scalable, energy-efficient, and robust IoT systems.